\documentclass{aastex62}

\graphicspath{{./}{figures/}}

\usepackage{lineno}
\usepackage{xspace}
\usepackage{amsmath}
\usepackage{layouts}

\usepackage{savesym}
\savesymbol{tablenum}
\usepackage{siunitx}
\restoresymbol{SIX}{tablenum}

\newcommand{\Abacus}{\textsc{Abacus}\xspace}

\newcommand{\kny}{\ensuremath{k_\mathrm{Nyquist}}}
\newcommand{\kmax}{\ensuremath{k_\mathrm{max}}}
\newcommand{\zinit}{\ensuremath{z_\mathrm{init}}}

\DeclareSIUnit \hkpc {\ensuremath{\mathit{h}^{-1}} \mathrm{kpc}}
\DeclareSIUnit \hMsun {\ensuremath{\mathit{h}^{-1}} M_\odot}

\newcommand{\bfq}{\mathbf{q}}
\newcommand{\bfx}{\mathbf{x}}
\newcommand{\bfv}{\mathbf{v}}
\newcommand{\bfs}{\mathbf{s}}
\newcommand{\bfk}{\mathbf{k}}
\newcommand{\bfw}{\mathbf{w}}

\makeatletter
\newcommand{\LHG}[1]{%
  \@bsphack
  \@esphack
}
\makeatother

\shorttitle{Halo Catalogs for Blind Challenges}
\shortauthors{Garrison \& Eisenstein}

\begin{document}

\title{Generating Approximate Halo Catalogs for Blind Challenges in Precision Cosmology}

\correspondingauthor{Lehman Garrison}
\email{lgarrison@cfa.harvard.edu}

\author{Lehman H.~Garrison}
\affil{Harvard-Smithsonian Center for Astrophysics, 60 Garden St., Cambridge, MA 02138}
%\nocollaboration

\author{Daniel J.~Eisenstein}
\affil{Harvard-Smithsonian Center for Astrophysics, 60 Garden St., Cambridge, MA 02138}

\begin{abstract}
We present a method for generating suites of dark-matter halo catalogs with only a few $N$-body simulations, focusing on making small changes to the underlying cosmology of a simulation with high precision.  In the context of blind challenges, this allows us to reuse a simulation by giving it a new cosmology after the original cosmology is revealed.  Starting with full $N$-body realizations of an original cosmology and a target cosmology, we fit a transfer function that displaces halos in the original so that the galaxy/HOD power spectrum matches that of the target cosmology.  This measured transfer function can then be applied to a new realization of the original cosmology to create a new realization of the target cosmology.  For a 1\% change in $\sigma_8$, we achieve 0.1\% accuracy to $k = 1h\,\mathrm{Mpc}^{-1}$ in the real-space power spectrum; this degrades to 0.3\% when the transfer function is applied to a new realization.  We achieve similar accuracy in the redshift-space monopole and quadrupole.  In all cases, the result is better than the sample variance of our $1.1h^{-1}\,\mathrm{Gpc}$ simulation boxes.
\end{abstract}

%% Keywords should appear after the \end{abstract} command. 
%% See the online documentation for the full list of available subject
%% keywords and the rules for their use.

\section{Introduction} \label{sec:intro}
Cosmological $N$-body simulations are a computationally expensive but important tool for forward-modeling a cosmological model to an observed distribution of galaxies.  Analytic methods cannot yet reproduce small-scale features in the galaxy field to the precision that observations provide, but $N$-body simulations are too expensive to densely sample the allowed cosmological parameter space.  Thus, much attention has turned to semi-numerical methods based on sparsely sampling the parameter space with $N$-body simulations.  Such techniques include ``emulation'', or interpolation, and ``warping'', or modifying the cosmology of a simulation output (e.g.~halo catalog).

Emulation typically focuses on certain key one-dimensional statistics like the power spectrum \citep[e.g.][]{Heitmann+2016}, 2PCF \citep[e.g.][]{Zhai+2018}, and halo mass function \citep[e.g.][]{McClintock+2018}.  This allows for fast evaluation in the space of cosmological parameters and possibly galaxy bias parameters, but analysis is limited to those statistics and to the prescribed galaxy bias models.  Warping, on the other hand, is typically slower but produces a full simulation output \citep[e.g.][]{Angulo_White_2010} or halo catalog \citep[e.g.][]{Mead_Peacock_2014} to which any catalog-based analysis may be applied.  This is our approach in this work.

Our motivation is to develop a warping technique suitable for use in blind mock challenges.  This is an analysis verification methodology in which an analyzer is given a mock galaxy catalog and asked to infer the underlying cosmology without knowing the true values.  This mimics real data analysis and helps avoid human bias in tuning of fitting parameters that could cause underestimation of the systematic error budget of the survey.  A fast warping technique is desirable for blind challenges since it allows a simulation to be used more than once.  Normally, a simulation can never be reused once its cosmology is revealed, but an accurate warping technique effectively re-blinds the simulation by changing the underlying cosmology.  This means that simulations can be reused between blind challenge epochs, so computational effort can be spent on making a few large-volume, high-quality catalogs instead of suites of single-use simulations.

%and multiple epochs are always necessary since analyzers will need to tune fitting parameters and fix bugs in their analysis pipelines once they see the unblinded results.

This blind-challenge context has implications for what kinds of rescaling techniques are allowable.  The analyzer should not be given any hints as to the direction of the change in cosmology, so changing the box size (e.g.~to hold the non-linear mass scale fixed) is not allowed.  Also, the precision of the mocks must be high enough that errors of the size of the analysis error budget can be identified, which itself is a fraction of the total survey error budget.

In this work, we present a new warping technique focusing on high precision ($\sim$0.1\%) for small changes in cosmology ($\sim$1\%), rather than rough precision for a large change in cosmology.  A 1\% change is sufficient to re-blind a catalog at the expected level of precision of analysis of a upcoming galaxy surveys.

To achieve this level of precision, our warping technique requires a full $N$-body realization of the target cosmology.  This one realization can be leveraged into many warped catalogs using more realizations of the original cosmology, as we will show.  Having a realization of the target cosmology is a requirement not all warping methods share, but it allows us to generate mocks of high precision.  Future extensions to this work will relax this requirement.

%start with full $N$-body realizations of the original and target cosmologies and fit a transfer function between the two (as explained in the body of this work).  This transfer function can then be applied to any new realization of the original cosmology with different phases. For example, our Abacus Cosmos suite of simulations \citep{Garrison+2018} has 20 realizations of the fiducial cosmology and 40 cosmologies with matched phases, so this technique can generate 800 mock catalogs from 40 transfer function measurements.

%Our technique uses large-scale mode replacement, as is common in other methods, and then a fitting step of a non-linear transfer function to displace halos .  We also rescale halo properties

\section{Warping}
\subsection{Outline}
Our warping procedure starts with two halo catalogs from phase-matched simulations with slightly different cosmologies: the ``original'' and the ``target''.  The catalogs include particle subsamples.  The goal is to modify the clustering of the original catalog to match that of the target.  The halo clustering does not simply following the mass clustering---we find the halo bias changes with cosmology at a non-negligible level---so we adjust the halo clustering directly by displacing halos and halo particles (see \autoref{sec:eulerian} for why we use displacements instead of re-labeling of halo mass).  The warping procedure consists of four main steps:
\begin{enumerate}
\item displace halos by the difference of the 2LPT initial conditions in the two simulations, scaled to the catalog redshift;
\item rescale halo radii (and other properties) as a function of abundance;
\item iteratively fit a transfer function that displaces the original halos to match the target halo power spectrum;
\item fit a second transfer function that adjusts the velocities of the original halos to match the target redshift-space halo power spectrum multipoles.
\end{enumerate}
Each of these steps will be explained in detail in the following sub-sections.

Although the resulting transfer functions are fit to a particular pair of phase-matched simulations, they are applicable to a new realization of the original cosmology (as we show in \autoref{sec:transferability}).  This is how we generate many warped catalogs from a single realization of the target cosmology.  With our Abacus Cosmos suite of simulations \citep{Garrison+2018}, for example, we have 20 realizations of a fiducial Planck cosmology and 40 with different wCDM cosmologies but matched phases, so this technique can generate 800 mock catalogs from 40 transfer function measurements.

\subsection{Initial Condition Residuals}
As with most warping procedures \citep[e.g.][]{Angulo_White_2010}, our first step is to replace the original large-scale simulation modes with those of the target cosmology.  We generate initial conditions (ICs) at \zinit\ in the Zel'dovich Approximation \citep[ZA,][]{Zeldovich_1970} for both cosmologies using the \texttt{zeldovich-PLT} code of \cite{Garrison+2016} which includes particle discreteness corrections.  We label these comoving displacement fields $\bfq_1(\bfx_i)$ and $\bfq_1'(\bfx_i)$, where the prime indicates a quantity in the target cosmology, $\bfx_i$ is the Lagrangian coordinate of particle $i$, and $\bfq_1$ is the first-order (ZA) displacement.  The associated velocities are labeled by $\bfv_1$ (we will leave the argument $\bfx_i$ implicit in the following).  We also generate the second-order Lagrangian Perturbation Theory displacements (2LPT, labeled $\bfq_2$) with the \Abacus configuration-space method \citep{Garrison+2016}, which uses two force evaluations and reversal of particle displacements.  We store the second-order part separately from the first-order part so we can apply the correct redshift and cosmology scaling for each.

These scalings are as follows:
\begin{gather}
\bfq_1(z) = \left[\frac{D(z)}{D(\zinit)}\right]\bfq_1, \\
\bfq_2(z) = \left[\frac{D(z)}{D(\zinit)}\right]^2\left[\frac{\Omega_m(z)}{\Omega_m(\zinit)}\right]^{-1/143}\bfq_2, \label{eqn:second_order}\\
\end{gather}
where $z$ is the catalog redshift and $D(z)$ is the linear growth factor, which \Abacus computes via direct integration of the linear growth equation.  \Abacus does not compute second-order growth factors, so in \autoref{eqn:second_order} we instead use the scalings of \cite{Bernardeau+2001} for the $\Omega_m$ dependence in a flat cosmology.  When computing the redshift dependence of primed quantities, we use the target cosmology.

The sum of the 1st and 2nd order residual differences between the cosmologies forms the total residual:
\begin{equation}
\Delta\bfq = (\bfq'_1(z) - \bfq_1(z)) + (\bfq'_2(z) - \bfq_2(z)).
\end{equation}

The velocity scaling is similar but carries a dependence on the growth rate $f$:
\begin{gather}\label{eqn:ic_vel_resid1}
\bfv_1(z) = \left[\frac{D(z)}{D(\zinit)}\right]\left[\frac{f(z)}{f(\zinit)}\right]\bfv_1, \\
\bfv_2(z) = \left[\frac{D(z)}{D(\zinit)}\right]^2\left[\frac{\Omega_m(z)}{\Omega_m(\zinit)}\right]^{-1/143}\left[\frac{\Omega_m(z)}{\Omega_m(\zinit)}\right]^{6/11}\bfv_2, \label{eqn:vel_second_order}\\
\Delta\bfv = (\bfv'_1(z) - \bfv_1(z)) + (\bfv'_2(z) - \bfv_2(z)).\label{eqn:ic_vel_resid3}
\end{gather}
Again, \Abacus computes the linear growth rate $f$ but not the corresponding second order quantity, so we use the known $\Omega_m$ scaling in \autoref{eqn:vel_second_order}.  We store velocities as comoving redshift-space displacements (that is, the same units as the positions), so we avoid any explicit cosmology dependence via $H(z)$.

We apply these position and velocity residuals to halo $h$ by taking the mean residual over the halo subsample particles:
\begin{equation}
\Delta\bar\bfq_h = \frac{1}{N_h^\mathrm{ss}}\sum_{i \in h} \Delta\bfq(\bfx_i),
\end{equation}
and likewise for the velocities.  $N_h^\mathrm{ss}$ is the number of subsample particles in halo $h$; we typically use 10\% of all halo particles.  The subsample particles contain an ID number that encodes the particle's Lagrangian coordinate $\bfx_i$, and thus the residual displacement $\Delta\bfq(\bfx_i)$ can be found.  This averaging process samples the initial Lagrangian patch from which the halo forms and effectively smooths the displacement field.  We have also tried introducing an explicit smoothing scale, but it has very little effect on the final result; any difference gets absorbed by the transfer function (\autoref{sec:transfer}).

We apply the comoving halo displacement $\Delta\bar\bfq_h$ to the halo center and to all its subsample particles.  The particles are simply ``advected'' along with the halo center.  We will consider internal halo structure changes in the next section.

In addition to the 10\% halo particle subsample, we also have a 10\% sample of all particles at the catalog redshift $z$ which is outputted during halo finding.  We will use this as a uniform sampling of the late-time matter density field in a later step.  To keep this field consistent with the displaced halo field, we save the initial condition residuals for these particles.

\subsection{Halo Property Rescaling}\label{sec:halo_rescaling}
We will now consider a basic prescription for rescaling internal halo structure by matching halo population statistics between cosmologies.  First, we must select the populations to match.  One choice is to rank halos by mass in one cosmology and make a mass cut, then rank halos in the other cosmology by mass and make a number-density cut so the two catalogs have the same galaxy density.  In this work, we use a mass cut of 100 particles in the original cosmology, chosen so that every halo has at least a few subsample particles (from which the IC displacements were computed in the previous section).  The abundance match cutoff will likely fall in the middle of a mass bin in the target cosmology---that is, many halos will have exactly 100 particles---so there is no clear abundance ordering in the last bin.  We select a random sample of halos in this bin mainly to avoid any pathologies from correlations between catalog order and spatial order.

% We typically use $N\ge100$ (halos of at least 100 particles).  This ensures that every halo has at least a few subsample particles; future work will lower this cutoff.  The abundance cutoff will likely fall in the middle of a mass bin, so we randomly select halos from the last bin to avoid spatial non-uniformity.  This randomness in the last bin actually contributes a fair amount of large-scale noise; it causes most of the scatter at low-$k$.

To rescale halo radii, we bin the halos by abundance, compute the median halo radius in each bin, and take ratio between cosmologies.  For small changes in cosmology, we observe that this ratio changes monotonically with log-abundance and can be fit with a line or low-order function.  We then adjust particles radially in each halo according to the ratio from this fit.  The exact radius definition we use depends on the halo finder.  For the simple friends-of-friends halos \citep{Davis+1985} considered here, we use $r_{50}$, or the 50th percentile of the radial particle distribution.

When considering redshift-space distortions, we apply this same rescaling procedure to the halo velocity dispersion $\sigma_v$.  Future extensions will consider more complicated changes to halo structure, such as changes to concentration.

\subsection{Transfer Function}\label{sec:transfer}
The previous two steps of applying IC residuals and rescaling halo properties are designed to increase the agreement between simulations but will not achieve our target of 0.1\% precision.  On large scales, small cosmological parameter changes cause the halo bias to shift even for abundance-selected halo samples (\autoref{fig:warping}).  On smaller scales, the step of applying the IC residuals tilts the power spectrum due to the effective smoothing imposed by averaging over subsample particles.  A power spectrum fitting routine would likely interpret this tilt as a change in cosmology.

We would like to force the clustering to match; we do so in Fourier space by moving halos according to a transfer function $T(k)$.  The transfer function operates isotropically on a ``late-time displacement field'': we compute the gradient of the gravitational potential of the late-time matter field and treat the resulting vector field as displacements.  This is the same idea as the Zel'dovich Approximation for initial conditions, except we are operating on the late-time matter density.  The idea is that these late-time displacements will trace bulk flows of the matter and that by pushing the halos along these flows we can increase or decrease the clustering of the catalog.  To allow for changes in the shape of the power spectrum, not just the mean amplitude, we modulate the late-time displacement field by an isotropic function $T(k)$ instead of a scalar.

%See \autoref{fig:late_time_disp} for an illustration.

%\begin{figure}[p]
%\begin{center}
%\includegraphics[width=1.0\textwidth]{AbacusCosmos_1100box_planck_00-0_late-time_za_disp_depth20_L100.png}
%\vspace{-5mm}
%\caption{An illustration of the ``late-time displacement field'' (\autoref{eqn:tfer}) on a small sub-volume of an AbacusCosmos box ($20h^{-1}\,\mathrm{Mpc}$ projection).  Halos are moved by this displacement field to force their clustering to match that of a target cosmology (in a $k$-dependent manner to allow for changes in power spectrum shape).  The displacements trace gradients of the large-scale gravitational potential as computed from the matter density (shaded background field).  The density is illustrated here in units of $\log(\delta + 1)$; the cell coarseness is the same as that used on the full volume.  The confluence of displacements in the lower half of the sub-volume is due to a massive cluster.  See \autoref{sec:transfer} for more discussion.
%\label{fig:late_time_disp}}
%\end{center}
%\end{figure}

% This is similar to the idea that a small change in cosmology is equivalent to a slight redshift rescaling (citation?).

Specifically, the late-time displacement field $\bfs(\bfx)$ is computed as
\begin{gather}
\label{eqn:tfer_real}
\bfs(\bfx) = \mathcal{F}^{-1}\left[ \tilde{\bfs}(\bfk) \right] \\
\label{eqn:tfer}
\tilde{\bfs}(\bfk) = 
\begin{cases}
ik^{-2}T(k)\tilde{\delta}_m(\bfk)\bfk, & k \leq \kmax; \\
\mathbf{0}, & k > \kmax, \\
\end{cases}
\end{gather}
where $\tilde{\delta}_m(\bfk)$ is the Fourier transform of the matter density field from subsample particles at the catalog redshift.  We apply the initial condition residuals $\Delta\bfq$ to these particles before computing the density field.  The inverse Fourier transform is indicated by $\mathcal{F}^{-1}$.

Since $\tilde{\bfs}(\bfk)$ is computed with an FFT, $\bfs(\bfx)$ exists on a lattice.  We evaluate a halo's displacement using tri-linear interpolation from the lattice to the halo center.  As with the IC residuals, the halo particles are advected along with the halo center.

We now discuss how we use optimization to find $T(k)$.  We parametrize $T(k)$ as $N_\mathrm{bin}$ discrete segments linearly spaced from the fundamental mode $k_\mathrm{fund}$ to \kmax, the latter of which is set by our desired analysis range.  We label this discretized transfer function $T(k_i)$.  Our target clustering metric is a pseudo-HOD power spectrum computed by giving every halo center a weight of 1 and every halo subsample particle a weight of $w_s=0.007$; this choice yields a satellite fraction of about 25\%.  Using every subsample particle as a ``fractional satellite'' has the effect of reducing the shot noise relative to a single HOD realization.  We compute the resulting overdensity field with TSC mass assignment and deconvolve the window function upon computing the power \citep{Jing_2005}.  We seek to minimize the difference in this quantity between original and target cosmologies.  Specifically, we minimize the difference in the real-space monopole $P_0(k)$:
\begin{gather}\label{eqn:chi2}
\chi^2 = \sum_{k_i}^{k_\mathrm{max}} \frac{[P_0(k_i) - P'_0(k_i)]^2}{2\alpha\sigma'^2_{P0}(k_i)},\\
\label{eqn:monopole_variance}
\sigma'^2_{P0}(k_i) = 2P_0'^2(k_i),
\end{gather}
where we take $P_0(k_i)$ to be the monopole power in bin $i$ and $\sigma'^2_{P0}(k_i)$ to be the monopole variance in that bin.  As before, a prime indicates a quantity from the target simulation.  We drop $N_\mathrm{modes}^{-1}$ from the formal sample variance definition since these are phase-matched simulations, and also we only care about the monopole variance relative to the quadrupole variance (during fitting of the redshift-space velocity transfer function; see below).  $\alpha$ is a numerical ``fudge factor'' to rescale $\chi^2$ to a convenient range for convergence testing; we typically use $\alpha = 10^{-6}$.

The process of drifting the halos and applying TSC mass assignment is non-linear, so we use a non-linear numerical optimizer to minimize $\chi^2$ with respect to $T(k_i)$.  For $\kmax=1h\,\mathrm{Mpc}^{-1}$ and a $512^3$ FFT mesh, we typically use $N_\mathrm{bin} = 10$, so this is a 10-dimensional optimization problem.  We have tried Powell's method \citep{Powell_1964} and Nelder-Mead \citep{Nelder_Mead_1965} from the SciPy package \citep{Scipy}; both work well.  For a trial transfer function $\hat T(k_i)$, a single $\chi^2$ evaluation consists of the following steps.
\begin{enumerate}
\item Apply $\hat T(k_i)$ to the late-time displacements $\tilde{\bfs}(\bfk)$ (\autoref{eqn:tfer}); call this field $\hat{\tilde{\bfs}}(\bfk)$.

\item Take the inverse FFT:
\begin{equation}
\hat{\bfs}(\bfx) = \mathcal{F}^{-1}[\hat{\tilde{\bfs}}(\bfk)].
\end{equation}

\item Interpolate the displacements $\hat{\bfs}(\bfx)$ to halo centers using tri-linear interpolation.
\item Apply the halo center displacements to halos and halo subsample particles.
\item Compute the resulting power spectrum monopole $P_0(k_i)$.
\item Compute $\chi^2$ (\autoref{eqn:chi2}).
\end{enumerate}
Note that $\tilde{\bfs}(\bfk)$ can be precomputed from the matter density field and is not updated during iteration.

Executing the above 6 steps takes about 3 seconds with \SI{1.5e6} halos and \SI{5.5e7} halo particles on a 24-core machine.  Powell's method uses about 6 steps and 1400 $\chi^2$ evaluations for a total optimization time of 1.2 hours.  The resulting $P_0$ matches $P'_0$ to 0.1\% down to $\kmax$ (except for the sample-variance-dominated large scales); see \autoref{sec:results}.

The choice of $\kmax$ is set by our science goals: we want to modify the power spectrum over the range that we will reasonably analyze.  Another consideration is the power spectrum mesh size: we want $\kmax$ to be somewhat smaller than $\kny$ of the mesh to avoid the worst of the aliasing effects, but a larger mesh slows requires more memory and makes the optimization slower.

\subsection{Redshift Space: Residuals, Velocity Dispersion, and Transfer Function}
The discussion thus far has focused on the real-space power spectrum, but we would like to match the redshift-space monopole and quadrupole, too.  Thus, we must consider how to warp the velocities.  We will apply initial condition residuals, rescale halo velocity dispersion, and apply a transfer function, as we did with the positions.  The IC residuals and transfer function modify the two-halo ``Kaiser'' redshift-space distortions (RSD); the velocity dispersion modifies the one-halo ``Finger-of-God'' RSD.

Applying the velocities from the IC residuals to the halo centers and particles is straightforward; the prescription is already given in Eqs.~\ref{eqn:ic_vel_resid1}--\ref{eqn:ic_vel_resid3}.  We apply the halo center kick identically to all halo particles.

We then rescale halo particle velocities according by matching velocity dispersion $\sigma_v$ as a function of abundance against the target cosmology.  As with the radius rescaling, we find that the ratio of the medians is a slowly changing function of abundance and can be fit with a line.  The particle velocities are simply scaled by this fit to the ratio.

Before starting the velocity warping, we apply one other correction.  The late-time displacements $\bfs(\bfx)$ in \autoref{eqn:tfer_real} imply a unique velocity $\bfw(\bfx)$ due to their dynamical origin (just as displacements in Zel'dovich Approximation initial conditions have a unique velocity):
\begin{equation}
\bfw(\bfx) = \frac{f(z)H(z)}{H_0}\bfs(\bfx).
\end{equation}
The $f(z)H(z)$ factor is a statement of redshift $z$ dynamics and gives the comoving velocity at that redshift.  The $H_0^{-1}$ factor converts this to the comoving redshift-space displacement for a $z=0$ observer. \LHG{is this right?}  We use redshift-space displacement units for all velocities.

The velocity transfer function framework is largely the same as the displacement transfer framework, except that we use the line-of-sight late-time velocities instead of the 3D late-time displacements.  The density field from which the late-time velocities are computed and the halo positions to which they are interpolated remain in real-space, but the power spectrum is computed on the redshift-space quantities.  We pre-apply the peculiar velocities to the halo particles as redshift-space distortions, so the velocity transfer function just has to drift them by the same amount as the halo center.

The velocity transfer function is still an isotropic monopole, as any velocity modifications must be isotropic.  We choose a line of sight for the RSD, however, so we only apply the transfer function to the $z$ velocities for efficiency while fitting.  We operate in the flat-sky approximation in a periodic simulation box, but the final warped catalog has the velocity modification applied isotropically, so the redshift space distortions should be accurate in any direction as long as the $z$ axis is not special.

We adopt a $\chi^2$ that includes both the redshift-space monopole and quadrupole:
\begin{gather}\label{eqn:chi2_zspace}
\chi^2 = \sum_{k_i}^{k_\mathrm{max}} \frac{[P_0(k_i) - P'_0(k_i)]^2}{2\alpha\sigma'^2_{P0}(k_i)} + \frac{[P_2(k_i) - P'_2(k_i)]^2}{2\alpha\sigma'^2_{P2}(k_i)},\\
\sigma'^2_{P2}(k_i) = 10P'_0(k_i)^2,
\end{gather}
where we have used $\ell = 2$ in the multipole variance $2(2\ell + 1)P'_0(k_i)^2$.  As with the monopole variance (\autoref{eqn:monopole_variance}), we drop the formal $N_\mathrm{modes}^{-1}$ dependence from the quadrupole variance, since these are phase-matched simulations and the variance only matters relative to the monopole.

The fitting takes about 9 steps with 2300 function evaluations, although it is largely converged in half that number.  This takes about 2.3 hours on a 24-core machine.

\section{Results}\label{sec:results}
\subsection{Outline}
In the following, we test our warping procedure on the \texttt{AbacusCosmos\_1100box\_00-0} and \texttt{AbacusCosmos\_1100box\_01-0} simulations which are available on the AbacusCosmos website\footnote{\url{https://lgarrison.github.io/AbacusCosmos/}}.  These are two phase-matched simulations with $1440^3$ particles in $1100h^{-1}\,\mathrm{Mpc}$ boxes and are identical except for a 1\% change in $\sigma_8$---the former (the original) has $\sigma_8 = 0.83$, while the latter (the target) has $\sigma_8=0.8383$.  We focus on this simple cosmology change for now while we validate the basic premise of our warping.  We use friends-of-friends halo catalogs at $z=0.5$ and 10\% halo and field particle subsamples.  We use halos of 100 particles or more in the original cosmology, which corresponds to halo mass \SI{3.7e12}{\hMsun}.  There are \SI{1.5e6} of these halos in a $1.3h^{-3}\,\mathrm{Gpc}^3$ volume, for a halo density of $\SI{1.1e-3}{}h^{-3}\,\mathrm{Mpc}^3$.

First, we examine the effect of the IC residuals, halo radius rescaling, and the transfer function in real space, then in redshift space.  Finally, we check that a transfer function measured on one pair of simulations can be applied to another with similar accuracy.

\subsection{Real Space}
\begin{figure}[p]
\begin{center}
\includegraphics[width=1.0\textwidth]{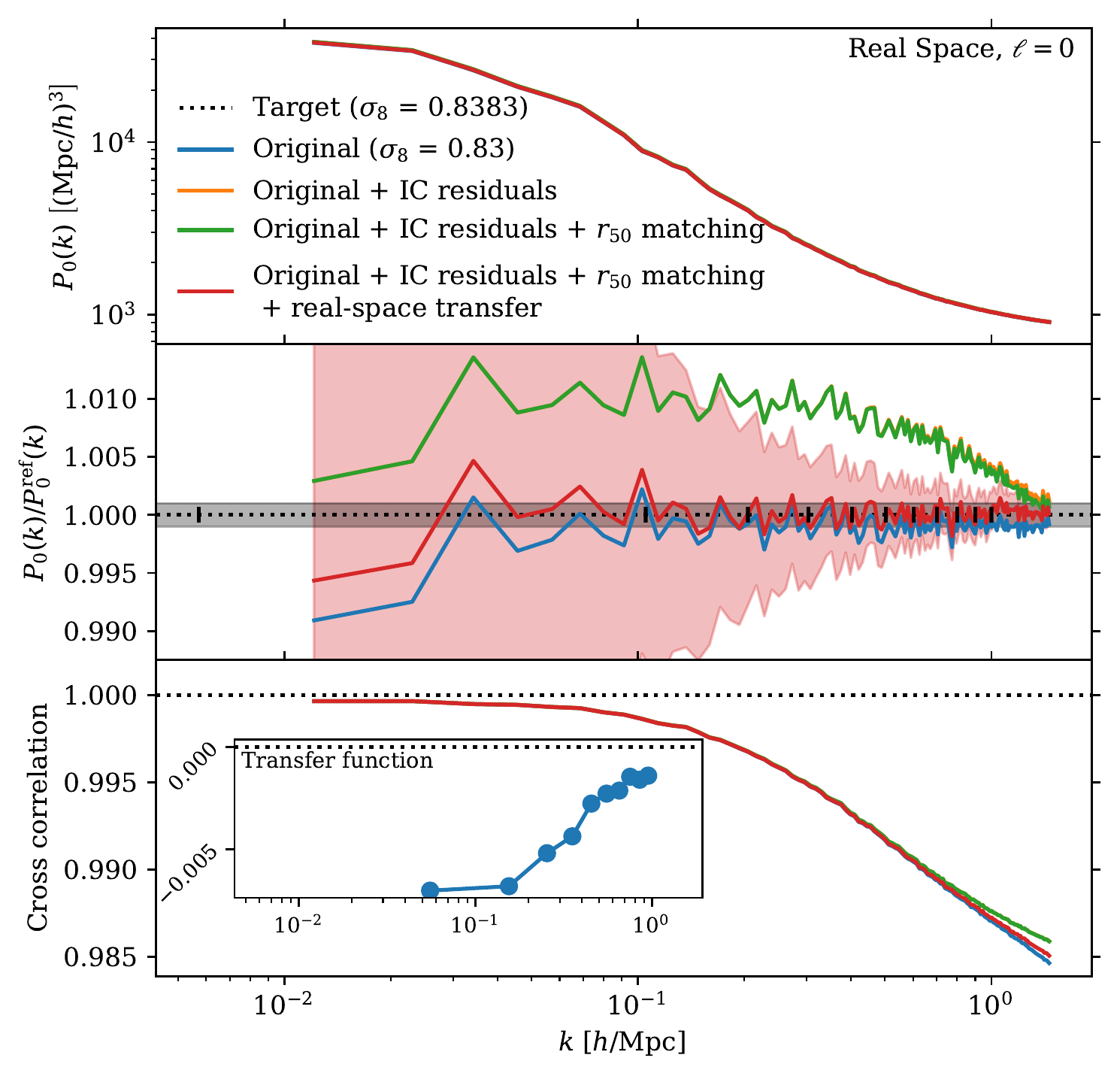}
%\vspace{-5mm}
\caption{The results of the real space warping in the real-space monopole at $z=0.5$.  The dotted lines are from the target simulation; each of the solid lines shows the original simulation during a stage of the warping.  Even with no modification, the original catalog matches the target catalog to a striking degree due to the halo bias changing to cancel the change in $\sigma_8$.  Thus, displacing halos by the difference in the ICs overshoots.  The final step of applying the transfer function brings the power into 0.1\% agreement with the target even past our chosen $\kmax = 1h\,\mathrm{Mpc}^{-1}$.  The broad shaded region indicates the sample variance error on the power spectrum; our fitting is consistently better than this variance.  The edges of the transfer function bins are marked with vertical ticks in the rectangular shaded region which indicates our target of 0.1\% precision.  The color scheme is matched to the redshift-space warping plots (Figures \ref{fig:warping_zspace} \& \ref{fig:warping_zspace_quadrupole}).
\label{fig:warping}}
\end{center}
\end{figure}

\autoref{fig:warping} shows the results of the warping procedure on the real-space power spectrum monopole and cross correlation.  First, we note that the halo bias evolves rapidly from the original to target simulations---for a 1\% change in $\sigma_8$, we should see a 2\% change in the halo power spectrum, assuming constant halo bias. (We do indeed measure exactly a 2\% change in the large-scale matter clustering.) Instead, we see almost no change; the halo bias has almost exactly canceled the change in $\sigma_8$.  Thus, when we apply the IC residuals from the large-scale modes, the halo power spectrum overshoots that of the target cosmology.

The bias cancellation does not mean that the catalog can be used as a realization of the new cosmology without any modification, of course.  This is seen most immediately in the redshift space $\ell=0$ and $\ell=2$ clustering (Figures~\ref{fig:warping_zspace} \& \ref{fig:warping_zspace_quadrupole}) where biases of 0.5--1.0\% are observed; our redshift-space warping will successfully deal with those in Section~\ref{sec:zspace_results}.

Despite the power spectrum overshoot due to applying the IC residuals, we see the small-scale cross correlation actually increases.  The cross correlation is already good on all scales---the error is smaller than 1.5\% for $k < 1h\,\mathrm{Mpc}^{-1}$---but this gives us confidence that the IC residuals are moving halos to more closely lie on top of their counterparts in the target cosmology.

\begin{figure}[p]
\begin{center}
\includegraphics[width=1.0\textwidth]{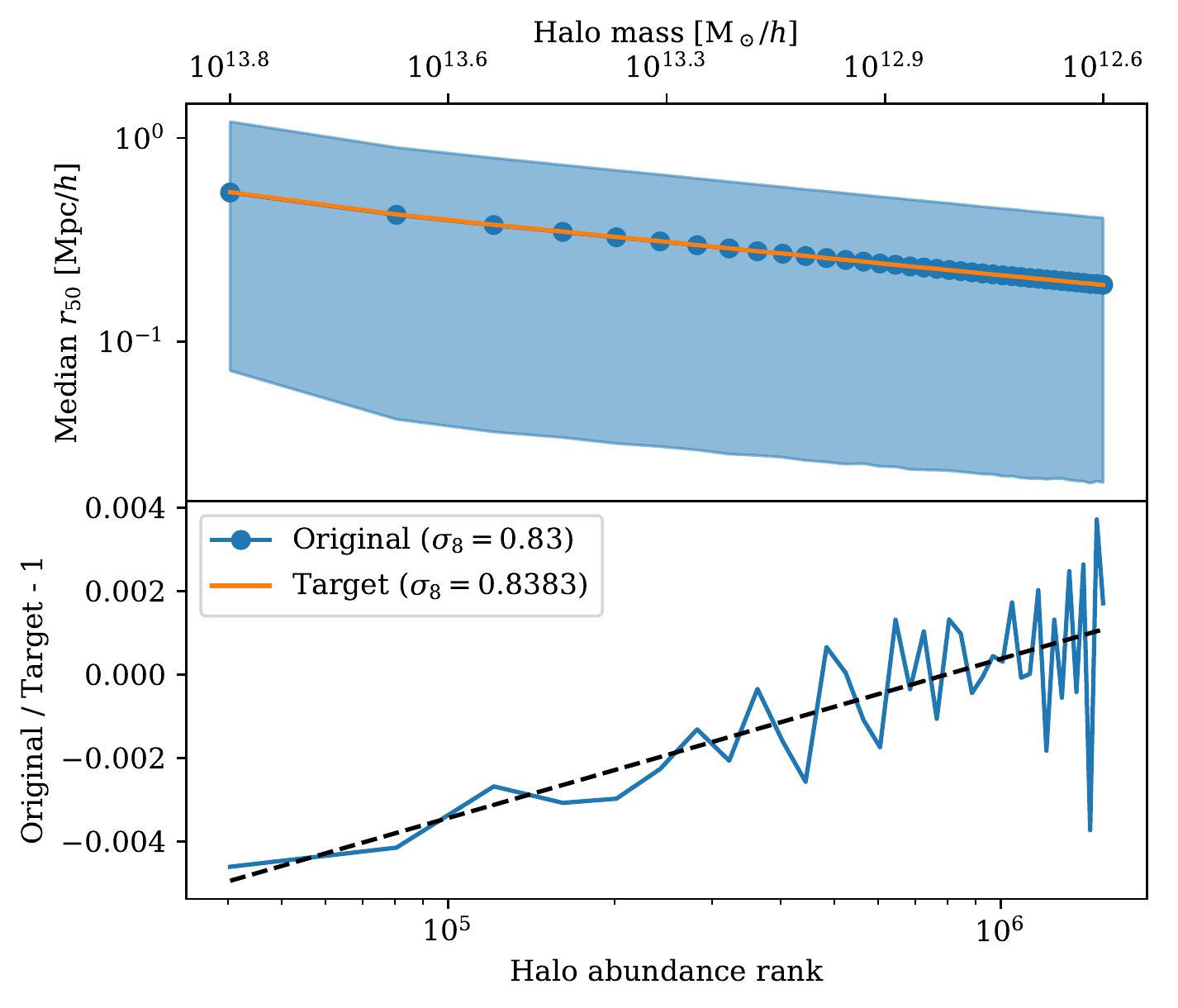}
%\vspace{-5mm}
\caption{\textit{Top panel:} median halo radius $r_{50}$ versus abundance in two cosmologies.  \textit{Bottom panel:} the fractional difference in the values in the top panel.  The relative radius changes monotonically with mass; we fit a line to the trend (bottom panel, dashed line) and move halo particles radially according to this fit.  The fact that this fit passes through zero indicates that massive halos are slightly larger and less massive halos are slightly smaller in the target cosmology.  The shaded region in the upper panel indicates the 25th to 75th percentiles of the $r_{50}$ distribution in each bin.  
\label{fig:r50}}
\end{center}
\end{figure}

The next step is to rescale halo radii according to $r_{50}$, the median halo radius, as a function of abundance.  The halo $r_{50}$ versus abundance for the two cosmologies is given in \autoref{fig:r50}.  This also shows the trend line that we fit to the change in median $r_{50}$; we move halo particles radially according to this fit.  The scatter in a given bin is quite large but the ratio of the medians appears robust.

\autoref{fig:warping} shows the effect of the radius rescaling on the power; as expected, the only impact is on small scales.  The radius changes are generally quite small (less than 0.5\%) which translates into a $<0.1\%$ shift in power even at our $\kmax = 1h\,\mathrm{Mpc}^{-1}$.

The direction of this change is such that agreement with the target simulation actually decreases.  This is because most halos in the original cosmology have larger $r_{50}$ than their counterparts in the target cosmology, despite the smaller $\sigma_8$.  This may be an effect of halos forming earlier and thus having higher concentrations with a higher $\sigma_8$.  This change is small enough that the cross-correlation is unaffected and the transfer function can easily compensate.  We prefer to match the 1-halo abundance statistics and deal with the consequences in the 2-halo clustering rather than neglect the former; otherwise, we are likely to end up with a catalog that looks accurate for this particular halo mass cut/HOD but not any other.

Finally, we fit the real-space transfer function using 6 steps of Powell's method (\autoref{sec:transfer}).  This brings the monopole power of the warped simulation into 0.1\% agreement with the target to $\kmax = 1h\,\mathrm{Mpc}^{-1}$; this is more precise than the sample variance of the box across the whole fitted range.

We note that the measured transfer function is a relatively smooth function of $k$ and would thus likely be amenable to parametrization as a low-dimensional function instead of 10 independent segments.  This would greatly accelerate convergence of the non-linear fit and probably eliminate the minor ``sawtooth'' effect observed at high $k$ due to use of a constant value to inside each bin.

We observe a tiny decrease in the cross correlation as a result of the transfer function.  Since the two-point clustering is the relevant quantity for most cosmology analysis, this very small loss of cross correlation is relatively unconcerning.  This is most likely an effect of halo mergers and splits that are not explicitly modeled here; such a problem would be compounded by our use of FoF halos, which are known to over-merge \citep[e.g.][]{More+2011}.  We will investigate this in future work with more sophisticated halo catalogs.

\subsection{Redshift Space}\label{sec:zspace_results}
\begin{figure}[p]
\begin{center}
\includegraphics[width=1.0\textwidth]{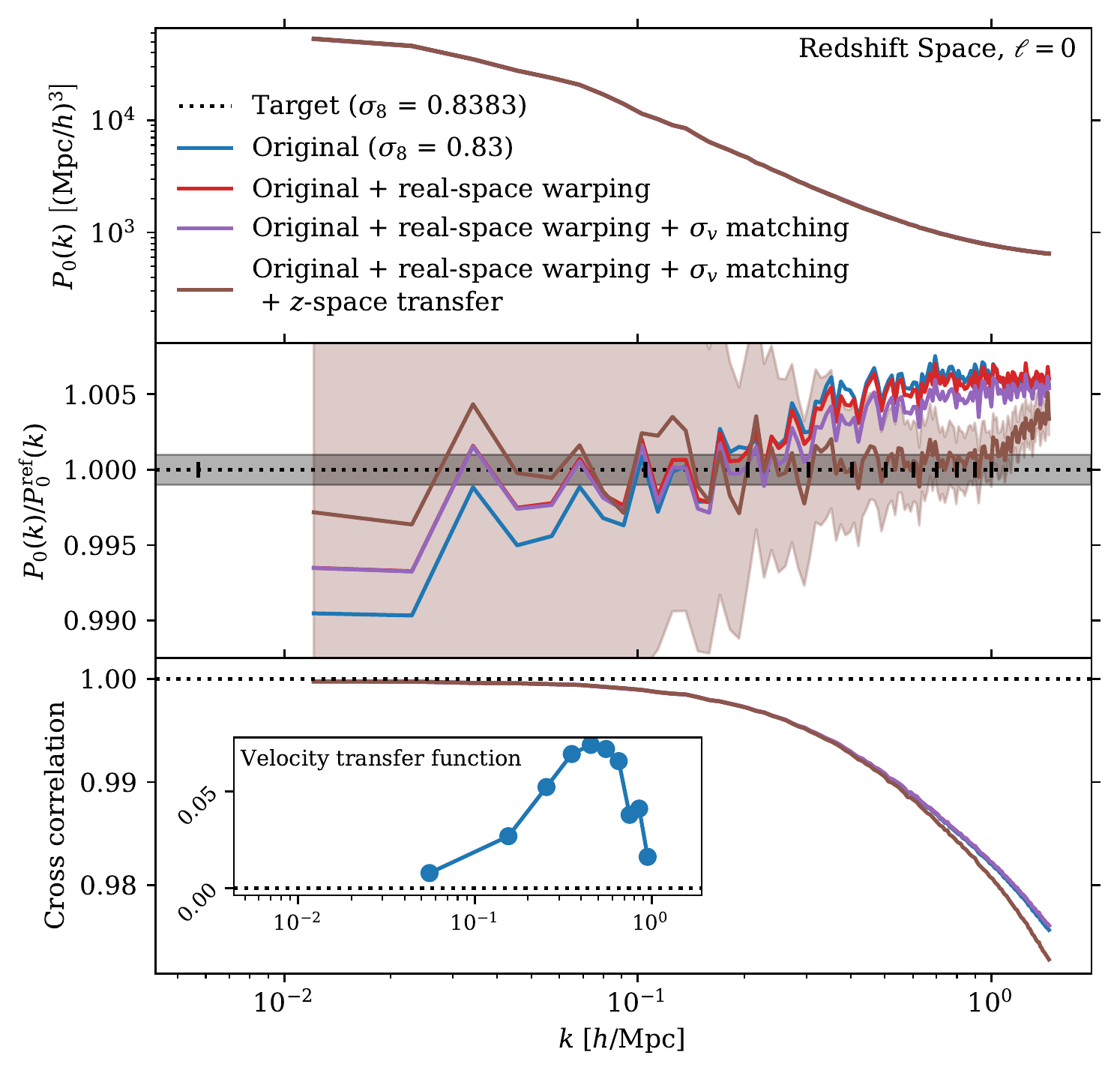}
%\vspace{-5mm}
\caption{The results of the warping on the redshift-space power spectrum monopole.  The real-space warping (i.e.~the end result of \autoref{fig:warping}) helps bring the large-scale power into alignment but fails on small scales.  Matching halo velocity dispersions and fitting a velocity transfer function match the power to the target within the sample variance to our chosen $\kmax=1h\,\mathrm{Mpc}^{-1}$.  See \autoref{sec:zspace_results}.  The color scheme is matched to the other warping plots.
\label{fig:warping_zspace}}
\end{center}
\end{figure}

\begin{figure}[p]
\begin{center}
\includegraphics[width=1.0\textwidth]{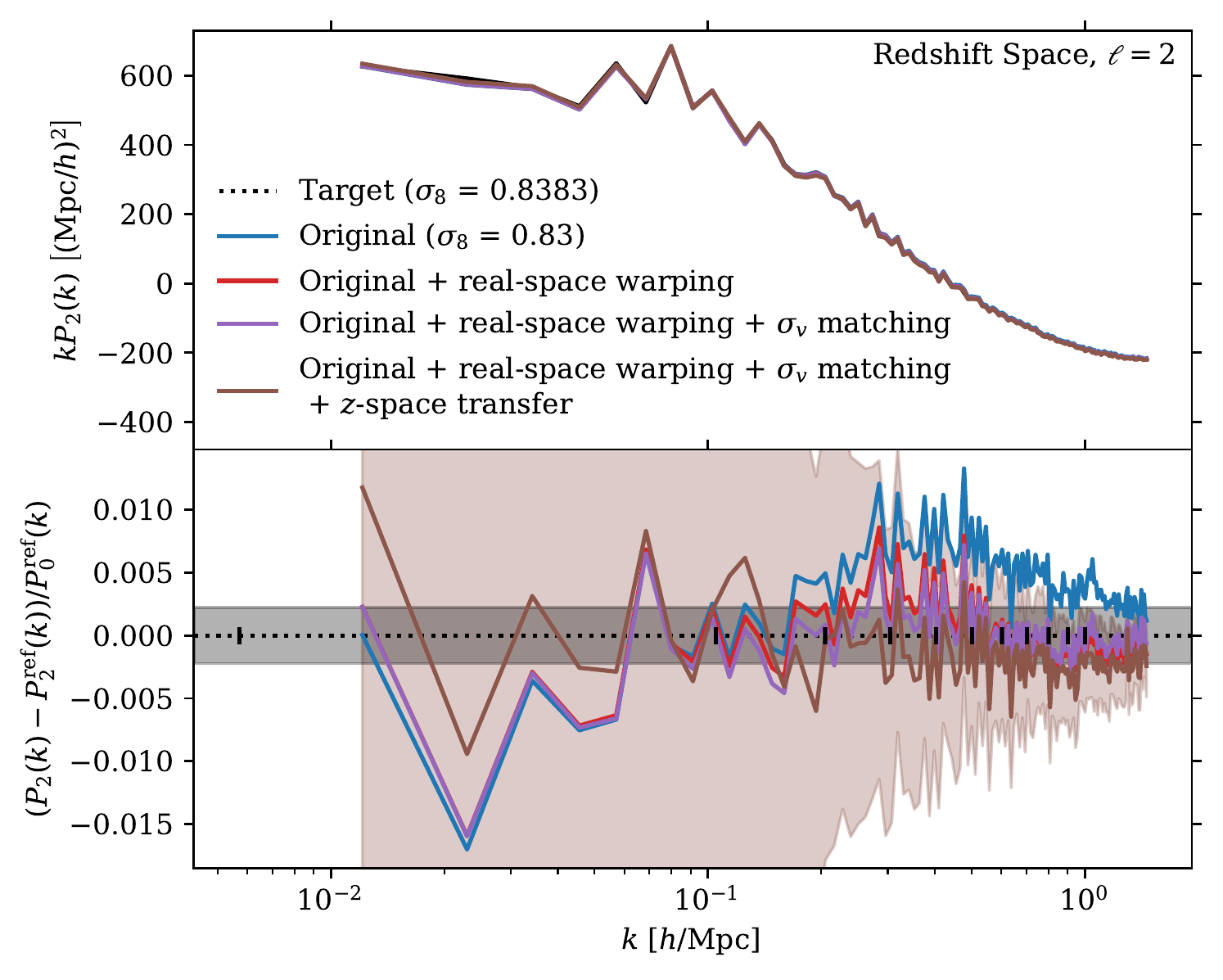}
%\vspace{-5mm}
\caption{The same results for warping in redshift-space as \autoref{fig:warping_zspace} but seen in the quadrupole.  Note that the second panel is the difference in the quadrupole relative to the monopole, due to the zero crossing of the quadrupole power around $k=0.2h\,\mathrm{Mpc}^{-1}$.  The horizontal shaded bar thus highlights the $\sqrt{5}\times0.1\%$ error region, since the quadrupole variance is 5 times that of the monopole.  Overall, these results show much the same trends as the redshift-space monopole which are that the real-space warping helps but the $\sigma_v$ matching and velocity transfer are necessary for precision matching.  The color scheme is matched to the other warping plots.
\label{fig:warping_zspace_quadrupole}}
\end{center}
\end{figure}

\begin{figure}[p]
\begin{center}
\includegraphics[width=1.0\textwidth]{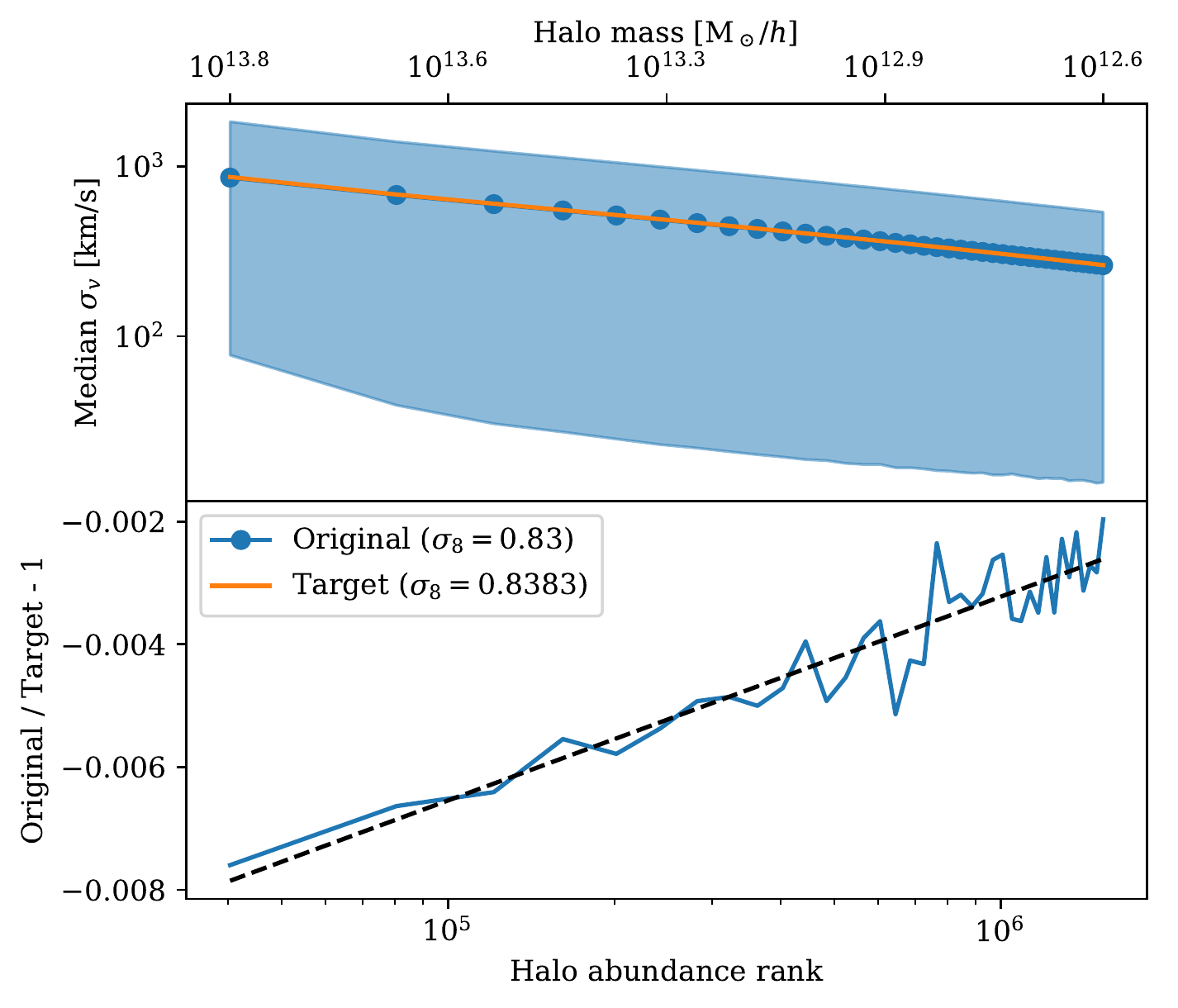}
%\vspace{-5mm}
\caption{\textit{Top panel:} median halo velocity dispersion $\sigma_v$ versus abundance in two cosmologies. \textit{Bottom panel:} the fractional difference in the values in the top panel.  As with the radius (\autoref{fig:r50}), the relative $\sigma_v$ is well fit by a line (bottom panel, dashed).  Unlike the radius, however, the sign of the difference is the same across the whole abundance range.  Thus, $\sigma_v$ is always larger in the target cosmology for the mass range examined here.  The shaded region in the upper panel indicates the 25th to 75th percentiles of the $\sigma_v$ distribution in each bin.
\label{fig:sigma_v}}
\end{center}
\end{figure}

Figures \ref{fig:warping_zspace} \& \autoref{fig:warping_zspace_quadrupole} show the results of the velocity warping procedure on the redshift-space monopole and quadrupole, respectively.  We see in both that the real-space warping restores some power on large scales, but the clustering amplitude is offset on small scales.  The first step of the velocity warping, matching $\sigma_v$, increases both monopole and quadrupole agreement.  In \autoref{fig:sigma_v}, we can see that the velocity dispersions are higher in the target cosmology, so boosting the dispersions in the original cosmology will tend to decrease the monopole and increase the quadrupole as galaxies are spread out along the line of sight.

The next step of the redshift-space warping is to fit the velocity transfer function, but unlike in real space, this transfer function must simultaneously satisfy the monopole and quadrupole.  Even under with this constraint, the transfer function does a very good job: the power matches the target within the sample variance of both multipoles to our chosen $\kmax=1.0h\,\mathrm{Mpc}^{-1}$.  The quadrupole is inherently nosier than the monopole, but it is clear that the fit is bringing the simulations into better agreement on all but the smallest scales.

The slight negative offset in the small-scale quadrupole pairs with the slight positive offset in the small-scale monopole.  This likely indicates tension between the two, which is unsurprising given that the velocity transfer function strictly operates on two-halo velocities and can make no modifications to internal halo structure.  On small scales, the power is dominated by internal halo velocity dispersions, so attempting to reproduce this with bulk halo motions is likely to fail.  This is likely the cause of the slight loss of cross-correlation on small scales as well.

\subsection{Transferring the Transfer Function}\label{sec:transferability}
\begin{figure}[p]
\begin{center}
\includegraphics[width=1.0\textwidth]{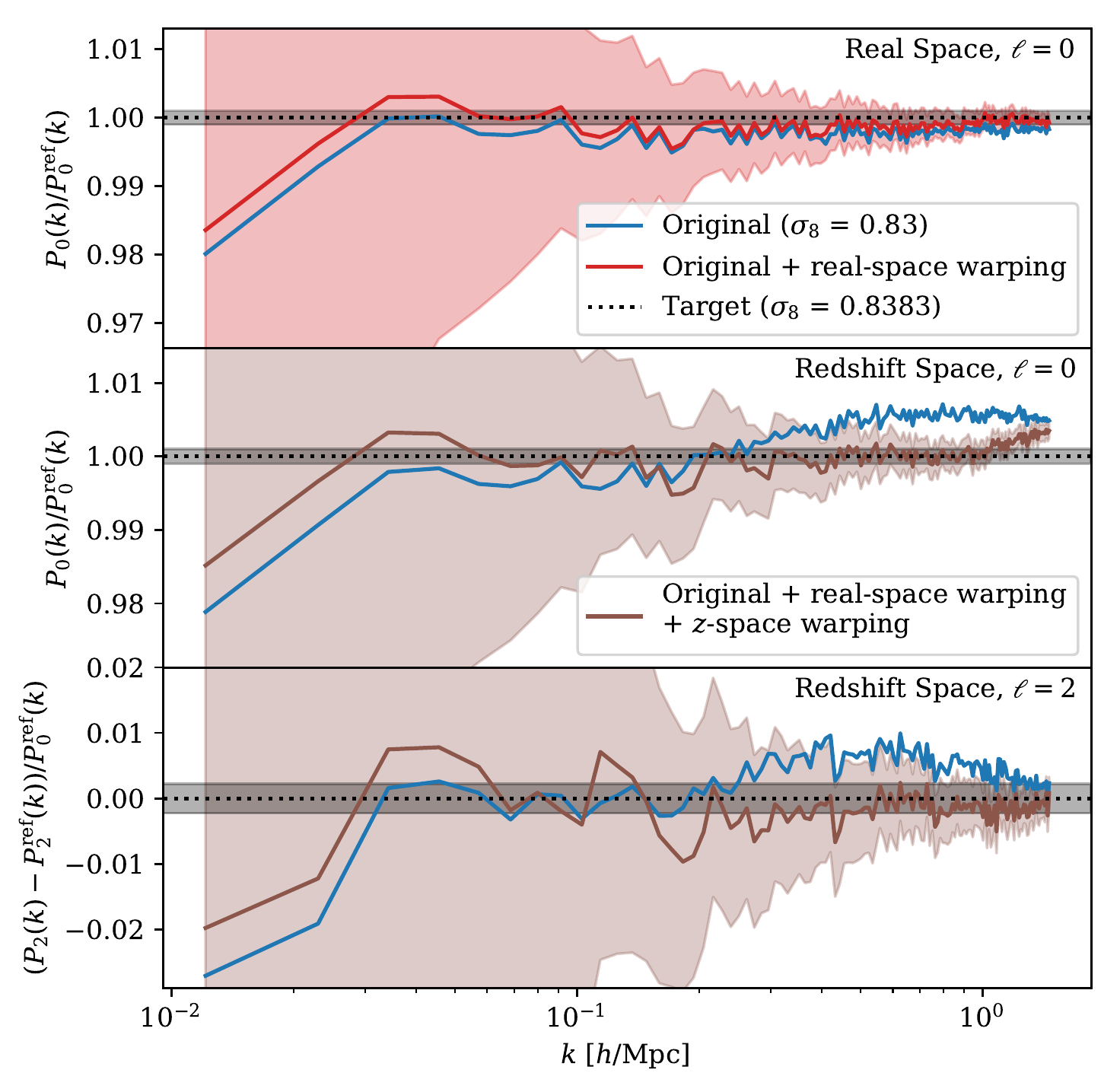}
%\vspace{-5mm}
\caption{A summary of the results of taking the real-space and velocity transfer functions measured on one pair of simulations and applying them to another pair with different initial phases.  The top panel shows the ratio of the real-space monopole with the target; the warped result is better than sample variance across the whole $k$ range.  The middle panel shows the redshift-space monopole, which is similarly accurate up to our chosen $\kmax = 1h\,\mathrm{Mpc}^{-1}$.  The bottom panel shows the difference of the redshift-space quadrupole relative to the monopole (due to the zero-crossing in the quadrupole; see \autoref{fig:warping_zspace_quadrupole}).  The quadrupole is noiser but matches better than sample variance across the whole $k$ range.  The horizontal shaded bar in the first two panels shows our target of 0.1\%; in the last panel, this is $\sqrt{5}\times0.1\%$ since it is relative to the monopole.  The real-space transfer function measurement is shown in \autoref{fig:warping}; the redshift-space (velocity) function measurement is shown in Figures \ref{fig:warping_zspace} \& \ref{fig:warping_zspace_quadrupole}.
\label{fig:transferability}}
\end{center}
\end{figure}

The utility of this warping methodology is based on the assumption that a transfer function measured on one pair of phase-matched simulations can be applied to a simulation with different initial phases.  We now test that assumption.  \autoref{fig:transferability} shows the application of the real-space transfer function (\autoref{fig:warping}) and redshift-space (velocity) transfer functions (Figures \ref{fig:warping_zspace} \& \ref{fig:warping_zspace_quadrupole}) to a new simulation.  The results are very good: the real space monopole matches the target better than sample variance across the whole $k$ range and better than 0.3\% across almost all of the $k$ range.  The results are similar in the redshift-space monopole and quadrupole: a mild degradation compared to the original phases but better than sample variance limits to $\kmax$.

The degradation in the real-space monopole appears largely as smooth deflection of peak amplitude $\sim0.3\%$ from $k=0.1$--$1h\,\mathrm{Mpc}^{-1}$.  This may be due to different non-linear couplings of the large-scale modes to small-scale structure in this pair of simulations compared with the pair on which the transfer function was measured.  At all scales this effect is sub-dominant to sample variance, however.

\LHG{the deflections at kmin are still strange.  Tests have shown this may be due to randomness in selecting halos in the N=100 bin.  Maybe different halos = different \# of subsample particles = different mean density.  Would that matter?}

\section{Discussion and Future Directions}
We have developed a warping framework for changing the cosmology of a simulation, focusing on highly accurate warped catalogs for small changes in cosmology rather than roughly accurate catalogs for large changes in cosmology.  We have used a halo displacement technique to modulate the clustering amplitude and thus avoided difficulties associated with an Eulerian re-weighting scheme (\autoref{sec:eulerian}). We have shown that the real-space power spectrum monopole can be controlled to 0.1\% precision, or 0.3\% when measured on one pair of simulations and applied to another.  The redshift space monopole and quadrupole are noisier and but can still be controlled to about 0.3\%.  In all cases, this error is better than the sample variance limit for a $1.1h^{-1}\,\mathrm{Gpc}$ box.

This work has focused on a simple change in cosmology (a 1\% change in $\sigma_8$) whose effects on the power spectrum are readily understood.  Even in this simple case, we found rapidly evolving halo bias, hence the need for a directly-fit transfer function rather than one motivated from initial conditions or first principles.  The next step of this work will be to test the effect of more complicated cosmology changes that also modulate the shape of the power spectrum.

The parametrization of $T(k)$ as 10 independent bins was chosen to allow full flexibility in the shape of the transfer function.  The measured transfer functions in this work were generally smooth which should admit lower-dimensional parameterizations; this is a natural next step for this work.  This will significantly accelerate the optimization step.  However, we expect that more complicated changes to the cosmology will cause shape variations in the transfer function, so we must ensure that any simpler parametrization can capture these as well.

We used friends-of-friends halos here for their simplicity, but FoF has well-known pathologies that may be causing some mild difficulties in the halo property rescaling and small-scale transfer function.  We expect this method will be applicable to other halo finders, such as \textsc{Rockstar} \citep{Behroozi+2013}  or spherical overdensity, and will test this in future work.

Finally, we used a pseudo-HOD prescription when assigning galaxies to halo particles in order to suppress shot noise, but we must test that the warped catalog behaves the same as the target catalog for a range of real HODs and HOD parameters.

We believe this technique will be suitable for applications that require high-quality mocks for small variations in cosmology, such as blind mock challenges.  Once we validate our method on more complicated cosmology changes, we will be able to produce suites of simulations suitable for warping.  The design of such a suite might be similar to that of our Abacus Cosmos simulations, with 20 realizations of a fiducial cosmology and 40 phase-matched simulations with different cosmologies.  Such a design would allow us produce 800 mock catalogs from 40 transfer function measurements.  Furthermore, it may be possible to interpolate the transfer functions between cosmologies.  This will maximize the utility of simulations in blind challenges, as we will be able to re-blind the same simulation many times.

\acknowledgments
LHG would like to thank Harshil Kamdar for work on an earlier iteration of this project.  We would also like to thank Doug Ferrer, Phil Pinto, and Marc Metchnik as co-authors of \Abacus, and Nina Maksimova, Duan Yutong, and Sownak Bose for helpful discussions.  This work has been supported by grant AST-1313285 from the National Science Foundation and by grant DE-SC0013718 from the U.S. Department of Energy.  DJE is further supported as a Simons Foundation investigator.

%% Similar to \facility{}, there is the optional \software command to allow 
%% authors a place to specify which programs were used during the creation of 
%% the manuscript. Authors should list each code and include either a
%% citation or url to the code inside ()s when available.

%\software{astropy \citep{2013A&A...558A..33A},  
%          }

%% Appendix material should be preceded with a single \appendix command.
%% There should be a \section command for each appendix. Mark appendix
%% subsections with the same markup you use in the main body of the paper.

%% Each Appendix (indicated with \section) will be lettered A, B, C, etc.
%% The equation counter will reset when it encounters the \appendix
%% command and will number appendix equations (A1), (A2), etc. The
%% Figure and Table counter will not reset.

\appendix
\section{Eulerian Transfer: Failed Warping Procedure}\label{sec:eulerian}
Rather than a non-linear fit to a transfer function on the late-time displacements, we initially tried to fit a transfer function directly on the Eulerian pseudo-HOD density field.  This approach had the advantage of being linear and solvable via a least-squares minimization.  The goal was to find the best-fit $T(k)$ such that
\begin{equation}
\tilde\delta'(\bfk) = T(k)\tilde\delta(\bfk).
\end{equation}
As before, the prime indicates a quantity in the target cosmology.  This can be posed as a complex least-squares minimization of
\begin{equation}
\sum_i |\tilde\delta'_i -  T(k)\tilde\delta_i|^2
\end{equation}
in each bin of $k$.
Imposing the constraint that $T(k)$ must be real gives the solution
\begin{equation}
\hat T(k) = \frac{\sum_i \operatorname{Re}(\tilde{\delta'^*_i}\tilde\delta_i)}{\sum_i |\tilde\delta_i|^2},
\end{equation}
where the asterisk denotes complex conjugation.  The halos (or the halo occupation statistics) would then be re-weighted by this transfer function to reproduce the target density field by construction.

This had two problems.  The first was that any roll-off in cross-correlation towards high $k$ would manifest as a suppressed transfer function, as evidenced by the fact that $\hat T(k)$ can be rewritten as the product of two terms, the first of which is the cross-correlation:
\begin{equation}
\hat T(k) = \left[\frac{\sum_i \operatorname{Re}(\tilde{\delta'^*_i}\tilde\delta_i)}{\sum_i |\tilde\delta'_i| |\tilde\delta_i|}\right]
\left[\frac{\sum_i |\tilde\delta'_i| |\tilde\delta_i|}{\sum_i |\tilde\delta_i|^2}\right].
\end{equation}
A cross-correlation of less than 1 was observed (\autoref{fig:warping}) and suppressed the best-fit transfer function.  The second term is related to the relative mode amplitudes and can be greater than 1, but any noise will reduce the covariance of the amplitudes and thus suppress the ratio.  We observed this effect as well.  A least-squares fit on $|\delta|$ or $|\delta|^2$ instead of $\delta$ would suffer from this same amplitude cross-correlation effect.

The second and more severe problem was that the re-weighted density field had no real-space positivity constraint, as it was constructed in Fourier space.  Thus, applying the transfer function caused large regions of $\delta < -1$ in the simulation voids.  The problem was severe enough that clipping these values to $-1$ distorted the power spectrum.  Re-weighting halos with negative values was deemed too unphysical to be tenable.

As a result of these difficulties with the Eulerian transfer function, we developed the transfer formalism with displacements used in the rest of this work.

% Could add figure showing negative voids and/or amplitude cross correlations

%\bibliographystyle{yahapj}
\bibliography{references}

\end{document}